%% file: main.tex
\documentclass[conference]{IEEEtran}
\IEEEoverridecommandlockouts
\usepackage{cite}
\usepackage{amsmath,amssymb,amsfonts}
\usepackage{algorithmic}
\usepackage{soul} 
\usepackage{graphicx}
\usepackage[caption=false]{subfig}
\usepackage{textcomp}
\usepackage{xcolor}
\def\BibTeX{{\rm B\kern-.05em{\sc i\kern-.025em b}\kern-.08em
    T\kern-.1667em\lower.7ex\hbox{E}\kern-.125emX}}
\usepackage{amsmath}
\newcommand\norm[1]{\left\lVert#1\right\rVert}    
\usepackage[nolist]{acronym}
\input{commands}
\input{acronyms.tex}  
\usepackage{pgfplots}
\usepackage{tikz}
\usetikzlibrary{patterns}
\usepackage{pdfpages}
\pgfplotsset{compat=1.17}

\begin{document}

\title{Temporal Averaging LSTM-based Channel Estimation Scheme for IEEE 802.11p Standard
}

\author{\IEEEauthorblockA{Abdul Karim Gizzini\IEEEauthorrefmark{1},
Marwa Chafii\IEEEauthorrefmark{3},
Shahab Ehsanfar\IEEEauthorrefmark{2},
Raed~M.~Shubair\IEEEauthorrefmark{3}
}

\IEEEauthorblockA{\IEEEauthorrefmark{1}ETIS, UMR8051,
CY Cergy Paris Université, ENSEA, CNRS, France\\
\IEEEauthorrefmark{2} Professorship of Communications Engineering, Technische Universit\"at Chemnitz, Germany \\
\IEEEauthorrefmark{3} Department of Electrical and Computer Engineering, New York University (NYU), Abu Dhabi 129188, UAE \\
Email: abdulkarim.gizzini, shahab.ehsan-far@etit.tu-chemnitz.de, \{marwa.chafii, raed.shubair\}@nyu.edu }}

\maketitle

\input{abstract}
\input{introduction}
\input{soa_estimators}

\input{proposed_estimator}

\input{simulation_results}
\input{conclusions}
\input{appendixA}
\input{acknowledgment}
\bibliographystyle{IEEEtran}
\bibliography{ref}
\end{document}

%% file: commands.tex
 \usepackage{bm}

\newcommand{\ma}  [1]{ \bm{#1} } 

\newcommand{\Ex}[1]{\mathrm{E}\left[ #1\right]} 

\newcommand{\set} [1]{{\mathcal {#1}}} 
\newcommand{\Kon} {\set{K}_{\text{on}}} 

\newcommand{\RS} {\set{RS}} 
\newcommand{\URS} {\set{URS}} 
\newcommand{\Kp} {\set{K}_{\text{p}}} 
\newcommand{\Kd} {\set{K}_{\text{d}}} 



%% file: acronyms.tex
\begin{acronym}
    \acro{C-ITS}{cooperative intelligent transportation system}
    \acro{V2V}{vehicle-to-vehicle}
    \acro{V2I}{vehicle-to-infrastructure}
    \acro{LSTM}{long short term memory}
    \acro{TDL}{tapped delay line}
    \acro{TA}{temporal averaging}
    \acro{VTV}{vehicle-to-vehicle}
    \acro{AWGN}{additive white Gaussian noise}
    \acro{RTV}{roadside-to-vehicle}
    \acro{STA}{spectral temporal averaging}
    \acro{SNR}{signal-to-noise ratio}
    \acro{CDP}{constructed data pilots}
    \acro{TRFI}{time domain reliable test frequency domain interpolation}
    \acro{MMSE-VP}{minimum mean square error using virtual pilots}
    \acro{DL}{deep learning}
    \acro{DPA}{data-pilot aided}
    \acro{AE-DNN}{auto-encoder deep neural network}
    \acro{DNN}{deep neural networks} 
    \acro{ISI}{inter-symbol-interference}
    \acro{OFDM}{orthogonal frequency-division multiplexing}
    \acro{LS}{least square}
    \acro{RS}{reliable subcarriers}
    \acro{URS}{unreliable subcarriers}
    \acro{MMSE}{minimum mean squared error}
    \acro{NMSE}{normalized mean-squared error}
    \acro{SoA}{state-of-the-art}
	\acro{BER}{bit error rate}
	\acro{RSU}{road side unit}
	\acro{AE}{auto-encoder}
		\acro{CP}{cyclic prefix}
\acro{PLCP}{physical layer convergence protocol}
\acro{MSE}{mean squared error}
\end{acronym}

%% file: abstract.tex
\begin{abstract}
In vehicular communications, reliable channel estimation is critical for the system performance due to the doubly-dispersive nature of vehicular channels. IEEE 802.11p standard allocates insufficient pilots for accurate channel tracking. Consequently, conventional IEEE 802.11p estimators suffer from a considerable performance degradation, especially in high mobility scenarios. Recently, {\ac{DL}} techniques have been employed for IEEE 802.11p channel estimation. Nevertheless, these methods suffer either from performance degradation in very high mobility scenarios or from large computational complexity. In this paper, these limitations are solved using a \ac{LSTM}-based estimation. The proposed estimator employs an {\ac{LSTM}} unit to estimate the channel, followed by {\ac{TA}} processing as a noise alleviation technique. Moreover, the noise mitigation ratio is determined analytically, thus validating the {\ac{TA}} processing ability in improving the overall performance.
Simulation results reveal the performance superiority of the proposed schemes compared to the recently proposed {\ac{DL}}-based estimators, while recording a significant reduction in the computational complexity.
\end{abstract}


%% file: introduction.tex
\section{Introduction} \label{introduction}

Vehicular communication technologies~\cite{ref1} describe a set of communication models that can be employed by vehicles in different application contexts, resulting in a well organized network infrastructure. The main motivation behind such technologies is to facilitate several future smart city applications including road safety and autonomous driving. 

In general, wireless communications in vehicular environment encounter a critical reliability challenge due to the doubly selective nature of the vehicular channel that varies rapidly especially in high mobility scenarios. Moreover, a precisely-estimated channel is critical for the equalization, demodulation, and decoding operations performed at the receiver. Therefore, robust and accurate channel estimation plays a crucial role in determining the overall system performance.  
IEEE 802.11p standard is an international standard that defines vehicular communications specifications. However, IEEE 802.11p standard is based initially on the IEEE 802.11a standard that was proposed for indoor environments, where an insufficient number of pilots is allocated for channel estimation. As a result,  IEEE 802.11p  conventional  estimators are based on the {\ac{DPA}} estimation, where the demapped data subcarriers, besides pilot subcarriers are employed in the channel estimation. In order to improve the {\ac{DPA}}  estimation  performance,  {\ac{STA}}  estimator~\cite{ref_STA} applies averaging  in  both  the  time  and  the  frequency  domains  as  post processing  operations  to the {\ac{DPA}} estimated channel. STA estimator performs well in low {\ac{SNR}} region, however it suffers  from  a  significant  performance  degradation  in  high  {\ac{SNR}} region especially in high mobility vehicular scenarios. The {\ac{TRFI}}~\cite{ref_TRFI} estimator assumes high correlation between successive symbols, thus it employs frequency domain interpolation to improve the {\ac{DPA}} performance. {\ac{TRFI}} estimator outperforms the {\ac{STA}} estimator in high {\ac{SNR}} region, however, these conventional estimators suffer from a considerable performance degradation in high mobility scenarios.

Recently, the rapid advancements in {\ac{DL}} and their successful applications in several domains, have sparked significant interest to adopt {\ac{DL}} techniques for wireless communication applications including channel estimation. {\ac{DL}} techniques are characterized by robustness, low-complexity, and good generalization ability making their integration into communication systems beneficial. Motivated by these advantages, {\ac{DL}} algorithms have been used for IEEE 802.11p channel estimator, where the authors in~\cite{ref_STA_DNN} and~\cite{ref_TRFI_DNN} employ {\ac{DNN}} as a post processing unit after the {\ac{STA}} and {\ac{TRFI}} conventional estimators respectively. The simulation results have showed that {\ac{STA}}-{\ac{DNN}} and {\ac{TRFI}}-{\ac{DNN}} are able to significantly improve the performance, however, they suffer from an error floor in high mobility scenarios. Another different {\ac{DL}}-based approach has been proposed in~\cite{ref_LSTM_DNN_DPA}, where {\ac{LSTM}} unit followed by {\ac{DNN}} network are employed as a pre-processing modules for channel estimation and noise error compensation. After that, {\ac{DPA}} estimation is applied using the {\ac{DNN}} output. This {\ac{LSTM}}-{\ac{DNN}}-{\ac{DPA}} estimator, outperforms the recently proposed {\ac{STA}}-{\ac{DNN}} and {\ac{TRFI}}-{\ac{DNN}} estimators, but it suffers from a considerable computational complexity due to the employment of two consecutive {\ac{DL}} networks.

In order to overcome the high complexity of the {\ac{LSTM}}-{\ac{DNN}}-{\ac{DPA}} estimator, while improving the {\ac{BER}} and {\ac{NMSE}} performances, in this paper we propose an {\ac{LSTM}}-based estimator, where the channel is estimated using an {\ac{LSTM}} unit only. After that, {\ac{DPA}} estimation is applied using the {\ac{LSTM}} estimated channel. Finally, unlike the {\ac{LSTM}}-{\ac{DNN}}-{\ac{DPA}} estimator where {\ac{DNN}} network is used for noise elimination, in the proposed estimator, {\ac{TA}} processing is employed as a noise alleviation technique where the noise alleviation ratio is calculated analytically. We also provide a detailed computational complexity analysis and make a comparison between the recent related work on {\ac{LSTM}}-{\ac{DNN}}-{\ac{DPA}} estimator and our proposed {\ac{LSTM}}-{\ac{DPA}}-{\ac{TA}} estimator.

The remainder of this paper is organized as follows: in Section~\ref{soa_estimators}, the system model followed by the recently proposed {\ac{SoA}} {\ac{DL}}-based IEEE 802.11p estimators are provided. The proposed {\ac{LSTM}}-{\ac{DPA}}-{\ac{TA}}, as well as the {\ac{TA}} processing analytical derivation, are described in Section~\ref{proposed_estimator}. In Section~\ref{simulation_results}, simulation results and computational complexity analysis are presented where the performance of the proposed scheme is evaluated in terms of \ac{BER}. Finally, the paper is concluded in Section~\ref{conclusions}. 

%% file: soa_estimators.tex
\section{SoA DNN-Based Channel Estimators} \label{soa_estimators}

\subsection{System Model}

IEEE 802.11p standard employs \ac{OFDM} transmission scheme with ${K = 64}$ total subcarriers. ${K_{\text{on}} = 52}$ active subcarriers are used, and they are divided into ${K_{\text{d}} = 48}$ data subcarriers and ${K_{\text{p}} = 4}$ pilot subcarriers. The remaining ${K_{\text{n}} = 12}$ subcarriers are used as a guard band. Moreover, IEEE 802.11p frame structure consists mainly of three parts. The first part contains the preamble that is used at the receiver for signal detection, timing synchronization, and channel estimation. Second, the signal field carries the transmission parameters like the employed code rate, modulation order, and frame length. Finally, we have the \ac{OFDM} data symbols. A detailed discussion of the IEEE 802.p standard and all its features are presented in~\cite{ref_IEEE}.

\begin{figure*}[t]
	\setlength{\abovecaptionskip}{6pt plus 3pt minus 2pt}
	\centering
  \includegraphics[width=0.95\textwidth]{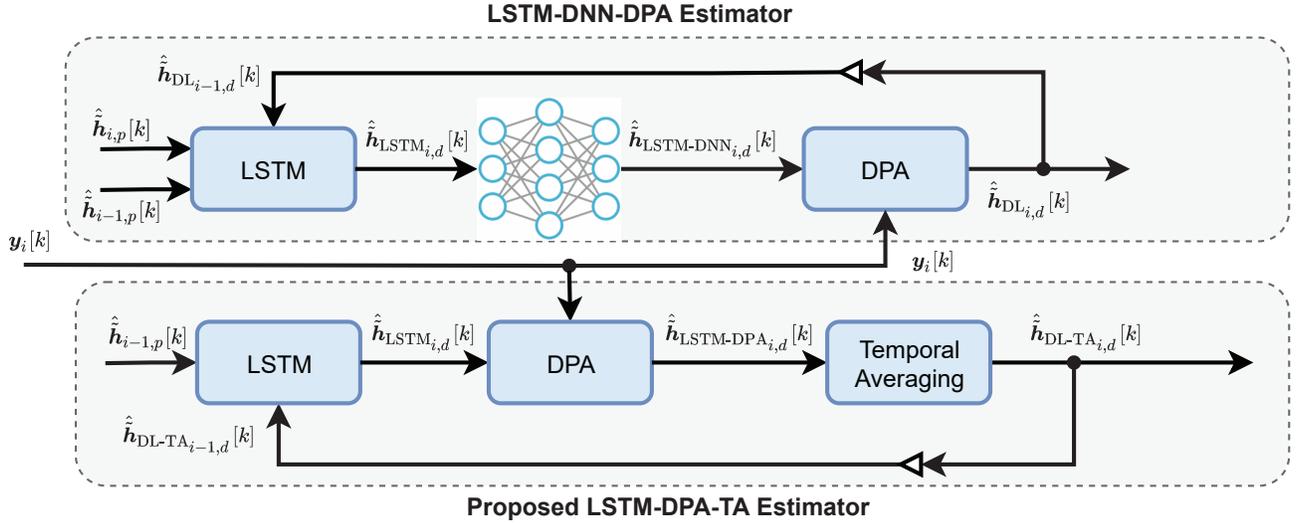}
  \caption{LSTM-based IEEE 802.11p channel estimators block diagram.}
  \label{fig:block_diagram}
\end{figure*}

In this paper, we assume perfect synchronization at the receiver, and we omit the signal field for simplicity. Therefore, we consider only the two long training symbols followed by $I$ {\ac{OFDM}} data symbols within each transmitted frame. The received {\ac{OFDM}} symbol can be expressed as: 
\begin{equation}
    \small
   \tilde{\ma{y}}_i[k] = \left\{
            \begin{array}{ll}        
                \tilde{\ma{h}}_{i,d}[k] \tilde{\ma{x}}_{i,d}[k] + \tilde{\ma{v}}_{i,d}[k],&\quad k \in \Kd \\
                \tilde{\ma{h}}_{i,p}[k] \tilde{\ma{x}}_{i,p}[k] + \tilde{\ma{v}}_{i,p}[k],&\quad k \in \Kp \\
            \end{array}\right. ,
\label{eq: xK}
\end{equation}
where $\tilde{\ma{x}}_{i}[k]$ and $\tilde{\ma{h}}_{i}[k]$ denote the $i$-th transmitted {\ac{OFDM}} symbol and its respective time variant frequency-domain channel response. Moreover, $\tilde{\ma{v}}_{i}[k]$ represents the frequency-domain counterpart of \ac{AWGN} of variance $\sigma^2$. $\Kd$ and $\Kp$ are the sets of data and pilot subcarriers indices respectively.    

\subsection{IEEE 802.11p Channel Estimators}

\subsubsection{DPA Basic Estimation}
The {\ac{DPA}} basic estimation employs the demapped data subcarriers besides pilots subcarriers to estimate the channel for the current {\ac{OFDM}} symbol, such that
\begin{equation}
\small
\ma{d}_i[k] =  \mathfrak{D} \big( \frac{\ma{y}_i[k]}{\hat{\tilde{\ma{h}}}_{\text{DPA}_{i-1}}[k]}\big)
,~ \hat{\tilde{\ma{h}}}_{\text{DPA}_{0}}[k] = \hat{\tilde{\ma{h}}}_{\text{LS}}[k],~ k \in \Kon, 
\label{eq: DPA_1}
\end{equation}
where $\mathfrak{D}(.)$ is the demapping operation to the nearest constellation point according to the employed modulation order. $\hat{\tilde{\ma{h}}}_{\text{LS}}[k]$ refers to the LS estimated channel at the received preambles denoted as $\ma{y}^{(p)}_{1}[k]$, and $\ma{y}^{(p)}_{2}[k]$, such that
\begin{equation}
\small
\hat{\tilde{\ma{h}}}_{\text{LS}}[k] = \frac{\ma{y}^{(p)}_{1}[k] + \ma{y}^{(p)}_{2}[k]}{2\ma{p}[k]},~k \in \Kon,
\label{eq: LS}
\end{equation}
where $\ma{p}[k]$ represents the frequency domain predefined preamble sequence. After that, the final {\ac{DPA}} channel estimated are updated as follows
\begin{equation}
\small
\hat{\tilde{\ma{h}}}_{\text{DPA}_{i}}[k] = \frac{\ma{y}_i[k]}{\ma{d}_i[k]},~ k \in \Kon.
\label{eq: DPA_2}
\end{equation}

It is worth mentioning that the {\ac{DPA}} basic estimation is considered as the starting point by most of the IEEE 802.11p channel estimators.


\subsubsection{STA-DNN Estimator}

The authors in \cite{ref_STA_DNN} discovered that employing an optimized {\ac{DNN}} after the conventional \ac{STA} estimator~\cite{ref_STA} leads to significant performance improvement, while recording lower computational complexity. This is because the conventional {\ac{STA}} estimation applies averaging in both frequency and time successively to the \ac{DPA} estimated channel in~\eqref{eq: DPA_2}, such that

\begin{equation}
\small
\hat{\tilde{\ma{h}}}_{\text{FD}_{i}}[k] = \sum_{\lambda = -\beta}^{\lambda = \beta} \omega_{\lambda} \hat{\tilde{\ma{h}}}_{\text{DPA}_{i}}[k + \lambda], ~ \omega_{\lambda} = \frac{1}{2\beta+1}, k \in \Kd,
\label{eq: STA_1}
\end{equation}

\begin{equation}
\small
\hat{\tilde{\ma{h}}}_{\text{STA}_{i}}[k] = (1 - \frac{1}{\alpha})  \hat{\tilde{\ma{h}}}_{\text{STA}_{i-1}}[k] + \frac{1}{\alpha}\hat{\tilde{\ma{h}}}_{\text{FD}_{i}}[k], k \in \Kon,
\label{eq: STA_2}
\end{equation}
where $\beta$ and $\omega_{\lambda}$ refer to the {\ac{STA}} frequency averaging window size and weight respectively, while $\alpha \geq 1$ defines the {\ac{STA}} time averaging weight. The main {\ac{STA}} limitation is that the averaging parameters should be updated according to the real-time channel statistics, that are not available in practice. Therefore, the {\ac{STA}} averaging parameters are fixed, resulting in a performance degradation especially in high \ac{SNR} regions and high mobility vehicular scenarios. However, as discussed in~\cite{ref_STA_DNN}, when the $\hat{\tilde{\ma{h}}}_{\text{STA}_{i}}[k]$ is fed as an input to an {\ac{STA}}-{\ac{DNN}} network, more time-frequency channel correlation can be captured, besides correcting the conventional STA estimation error. Thus, the overall performance is significantly improved. However, an error floor still appears in high mobility vehicular scenario, especially in high {\ac{SNR}} region. 

\subsubsection{TRFI-DNN Estimator}
The {\ac{DPA}} estimation in~\eqref{eq: DPA_2} can be further improved by applying the {\ac{TRFI}}~\cite{ref_TRFI}, where the subcarriers of the received {\ac{OFDM}} symbol are divided into: (\textit{i}) {$\RS_{i}$} set: that includes the reliable subcarriers indices,  and (\textit{ii}) {$\URS_{i}$} set: which contains the unreliable subcarriers indices. Then, the estimated channels for the {$\URS_{i}$} are interpolated using the {$\RS_{i}$} channel estimates. {\ac{TRFI}} employs frequency domain cubic interpolation assuming high correlation between two successive received {\ac{OFDM}} symbols. This procedure can be expressed as follows

\begin{itemize}
    \item Equalize the previously received {\ac{OFDM}} symbol by ${\hat{\tilde{\ma{h}}}_{\text{TRFI}_{i-1}}[k]}$ and ${\hat{\tilde{\ma{h}}}_{\text{DPA}_{i}}[k]}$, such that
    \begin{equation}
    \small
    \begin{split}
    {\ma{d^\prime}_{i-1}[k]} = \mathfrak{D} \big( \frac{\ma{y}_{i-1}[k]}{\hat{\tilde{\ma{h}}}_{\text{DPA}_{i}}[k]} \big), ~
    {\ma{d^{\prime\prime}}_{i-1}[k]} =  \mathfrak{D} \big( \frac{\ma{y}_{i-1}[k]}{\hat{\tilde{\ma{h}}}_{\text{TRFI}_{i-1}}[k]} \big). 
    \label{eq: TRFI_1}
    \end{split}
    \end{equation}

    \item According to the demapping results, the subcarriers are grouped as follows
    \begin{equation}
     \small
       \left\{
        \begin{array}{ll}
            \RS_{i} \leftarrow \RS_{i} + {k} ,&\quad \ma{d^\prime}_{i-1}[k] = \ma{d^{\prime\prime}}_{i-1}[k] \\
            \URS_{i} \leftarrow \URS_{i} + {k},&\quad \ma{d^\prime}_{i-1}[k] \neq \ma{d^{\prime\prime}}_{i-1}[k]
        \end{array}\right. .
       \label{eq: TRFI_2}
       \end{equation}
\item Finally, frequency domain cubic interpolation is employed to estimate the channels at the {$\URS_{i}$} as follows

\begin{equation}
     \small
      \hat{\tilde{\ma{h}}}_{\text{TRFI}_{i}}[k] = \left\{
        \begin{array}{ll}
            \hat{\tilde{\ma{h}}}_{\text{DPA}_{i}}[k] ,&\quad k \in \RS_{i} \\
            \text{Cubic Interpolation},&\quad k \in \URS_{i}
        \end{array}\right. .
       \label{eq: TRFI_3}
       \end{equation}
\end{itemize}

In order to outperform the {\ac{STA}}-{\ac{DNN}} performance limitation in high mobility vehicular scenarios (high \ac{SNR} region), the authors in~\cite{ref_TRFI_DNN}, used the same optimized {\ac{DNN}} architecture as in~\cite{ref_STA_DNN}, but with $\hat{\tilde{\ma{h}}}_{\text{TRFI}_{i}}[k]$ as an input instead of  $\hat{\tilde{\ma{h}}}_{\text{STA}_{i}}[k]$. \ac{TRFI}-\ac{DNN} corrects the cubic interpolation error, besides learning the channel frequency domain correlation, thus improving the performance in high {\ac{SNR}} region.

\subsubsection{LSTM-DNN-DPA Estimator}
Unlike the recently proposed \ac{DNN}-based estimators, where the {\ac{DL}} processing is employed after the conventional estimators, the work performed in~\cite{ref_LSTM_DNN_DPA} show that employing the {\ac{DL}} processing before the conventional estimator (specifically the {\ac{DPA}} estimation), could improve significantly the overall performance. In this context, the authors proposed to use two cascaded {\ac{LSTM}} and {\ac{DNN}} networks to estimate the channel for the current {\ac{OFDM}} symbol as shown in Fig.~\ref{fig:block_diagram}. After that, a {\ac{DPA}} estimation is applied using the {\ac{LSTM}}-{\ac{DNN}} estimated channel. Even though, the {\ac{LSTM}}-{\ac{DNN}}-{\ac{DPA}} estimator can outperform the recently proposed \ac{DNN}-based estimators, it suffers from a considerable computational complexity that rises from the employment of two {\ac{DL}} networks.

%% file: proposed_estimator.tex
\section{Proposed LSTM-based Estimation Scheme} \label{proposed_estimator}
In this section, the proposed LSTM based estimator is discussed. First of all, a brief description of {\ac{LSTM}} network is presented. After that, the proposed {\ac{TA}} noise power alleviation is analytically derived.

\subsection{LSTM Overview}

{\ac{LSTM}} networks are basically proposed to deal with sequential data where the order of the data matters and there exists a correlation between the previous and the future data. In this context, {\ac{LSTM}} networks are defined with an appropriate architecture that can learn the data correlation over time, thus giving the {\ac{LSTM}} network the ability to predict the future data based on the previous observations.

{\ac{LSTM}} unit contains computational blocks known as gates which are responsible for controlling and tracking the information flow over time. The {\ac{LSTM}} network mechanism can be explained in four major steps as follow
  \paragraph{Forget the irrelevant information} In general, the {\ac{LSTM}} unit classify the input data into relevant and irrelevant information. The first processing step is to eliminate the irrelevant information that are not important for the future data prediction. This can be performed through the forget gate that decides which information the {\ac{LSTM}} unit should keep, and which information it should delete. The forget gate processing is defined as below
    \begin{equation}
    \small
    {\ma{f}}_{t} = {\sigma} (\ma{W}_{f, t}\bar{\ma{x}}_{t} + \ma{W}^{\prime}_{f,t}\bar{\ma{h}}_{t-1} + \bar{\ma{b}}_{f,t}),
    \label{eq: lstm_fg}
    \end{equation}
    where ${\sigma}$ is the sigmoid function, $\ma{W}_{f,t} \in \mathbb{R}^{P \times K_{in}}$,  $\ma{W}^{\prime}_{f,t} \in \mathbb{R}^{P \times P}$ and $\bar{\ma{b}}_{f,t} \in \mathbb{R}^{P \times 1}$ are the forget gate weights and biases at time $t$, $\bar{\ma{x}}_{t} \in \mathbb{R}^{K_{in} \times 1}$ and $\bar{\ma{h}}_{t-1}$ denote the {\ac{LSTM}} unit input vector of size $K_{in}$, and the previous hidden state of size $P$ respectively.
\paragraph{Store the relevant new information} After classifying the relevant information, the {\ac{LSTM}} unit applies some computations on the selected information through the input gate
    \begin{equation}
    \small
    {\bar{\ma{i}}_{t}} = {\sigma} (\ma{W}_{\bar{\ma{i}}, t}\bar{\ma{x}}_{t} + \ma{W}^{\prime}_{\bar{\ma{i}},t}\bar{\ma{h}}_{t-1} + \bar{\ma{b}}_{\bar{\ma{i}},t}),
    \label{eq: lstm_ing}
    \end{equation}
    \begin{equation}
    \small
    {\tilde{{\ma{c}}}}_{t} = \text{tanh} (\ma{W}_{{\tilde{{\ma{c}}}}, t}\bar{\ma{x}}_{t} + \ma{W}^{\prime}_{{\tilde{{\ma{c}}}},t}\bar{\ma{h}}_{t-1} + \bar{\ma{b}}_{{\tilde{{\ma{c}}}},t}).
    \label{eq: lstm_incg}
    \end{equation}
    
\paragraph{Update the new cell state} Now, the {\ac{LSTM}} unit should update the current cell state ${{{\ma{c}}}}_{t}$ based on the two previous steps such that
    \begin{equation}
    \small
    {{{\ma{c}}}}_{t} = {\ma{f}}_{t} \odot {\ma{c}}_{t-1} +  \bar{\ma{i}}_{t} \odot {\tilde{{\ma{c}}}}_{t}, 
    \label{eq: lstm_cell_state}
    \end{equation}
    where $\odot$ denotes the Hadamard product. 
    \paragraph{Generate the LSTM unit output} The final processing step is to update the hidden state and generate the output by the output gate. The output is considered as a cell state filtered version and can be computed such that
    \begin{equation}
    \small
    {\ma{o}}_{t} = {\sigma} (\ma{W}_{o, t}\bar{\ma{x}}_{t} + \ma{W}^{\prime}_{o,t}\bar{\ma{h}}_{t-1} + \bar{\ma{b}}_{o,t}),
    \label{eq: lstm_og}
    \end{equation}
    \begin{equation}
    \small
    {\bar{{\ma{h}}}}_{t} =  {\ma{o}}_{t} \odot \text{tanh}{\ma{c}}_{t}.
    \label{eq: lstm_hidden_state}
    \end{equation}

We note that, in literature there exists several {\ac{LSTM}} architecture variants, where the interactions between the {\ac{LSTM}} unit gates are modified. The authors in~\cite{ref_lstm_var} performed a nice comparison of popular {\ac{LSTM}} architecture variants. However, for the proposed estimator we focus on the classical {\ac{LSTM}} unit architecture. 

\begin{figure}[t]
	\centering
  \includegraphics[width=\columnwidth]{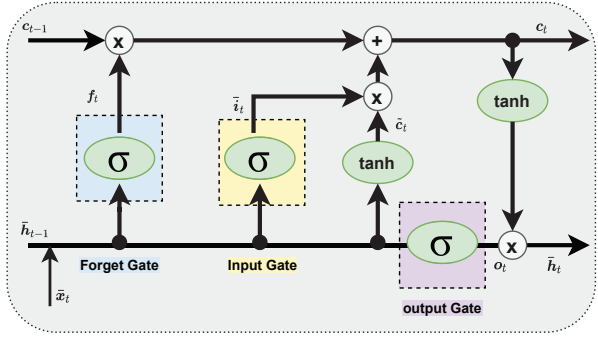}
  \caption{LSTM unit architecture.}
  \label{fig:lstm_archi}
\end{figure}

\subsection{Proposed LSTM-TA Estimator}

Unlike the {\ac{LSTM}}-{\ac{DNN}}-{\ac{DPA}} estimator, where the {\ac{LSTM}} and {\ac{DNN}} networks are used for channel estimation and noise compensation respectively, the proposed estimator employs only {\ac{LSTM}}. Moreover, the {\ac{LSTM}} input dimension is decreased to include only $K_{on}$ subcarriers, therefore the overall computational complexity is significantly reduced. The proposed estimator proceeds as follows:
\paragraph{LSTM-based prediction}: The first step is to estimate the channel for the current received {\ac{OFDM}} symbol employing the previous estimated channel $\hat{\tilde{\ma{h}}}_{\text{DL-TA}_{i-1,d}[k]}$,  where the  $i$-th {\ac{LSTM}} unit input is denoted by $\tilde{\bar{{\ma{x}}}}_{{i}} \in \mathbb{R}^{2 K_{on} \times 1}$, where
    \begin{equation}
         \small
         \bar{\ma{x}}_{i} = \left\{
        \begin{array}{ll}
            \hat{\tilde{\ma{h}}}_{\text{LSTM}_{i-1,d}}[k] ,&\quad k \in \Kd \\
            \hat{\tilde{\ma{h}}}_{{i-1,p}}[k],&\quad k \in \Kp
        \end{array}\right. .
       \label{eq: proposed2}
       \end{equation}
    
    $\tilde{\bar{{\ma{x}}}}_{{i}}$ is obtained by applying complex to real valued conversion to $\bar{\ma{x}}_{i}$ by stacking its real and imaginary values in one vector. We note that $\hat{\tilde{\ma{h}}}_{{i-1,p}}[k]$ denotes the {\ac{LS}} estimated  channel at the $\Kp$ subcarriers. After that, $\tilde{\bar{{\ma{x}}}}_{{i}}$ is processed by the {\ac{LSTM}} unit, such that
    \begin{equation}
        \small
        \hat{\tilde{\ma{h}}}_{\text{LSTM}_{i,d}} = \Omega_{\text{LSTM}}(\tilde{\bar{\ma{x}}}_{i},\Theta),
        \label{eq: proposed1}
        \end{equation}
        where $\Omega_{\text{LSTM}}$ is the {\ac{LSTM}} unit processing with overall weights denoted by $\Theta$.
\paragraph{DPA estimation} The {\ac{LSTM}} estimated channel undergoes {\ac{DPA}} estimation using the $i$-th received {\ac{OFDM}} symbol as follows
    \begin{equation}
    \small
    \ma{d}_{\text{LSTM}_{i}}[k] =  \mathfrak{D} \big( \frac{\ma{y}_i[k]}{\hat{\tilde{\ma{h}}}_{\text{LSTM}_{i-1}}[k]}\big)
    ,~ \hat{\tilde{\ma{h}}}_{\text{LSTM}_{0}}[k] = \hat{\tilde{\ma{h}}}_{\text{LS}}[k],
    \label{eq: proposed3}
    \end{equation}
    
    \begin{equation}
    \small
    \hat{\tilde{\ma{h}}}_{\text{LSTM-DPA}_{i}}[k] = \frac{\ma{y}_i[k]}{\ma{d}_{\text{LSTM}_{i}}[k]}.
    \label{eq: proposed4}
    \end{equation}
    
\paragraph{TA processing}  Finally, in order to alleviate the impact of the AWGN noise, {\ac{TA}} processing is applied to the $\hat{\tilde{\ma{h}}}_{\text{LSTM-DPA}_{i}}[k]$ estimated channel, such that
    \begin{equation}
    \small
    \hat{\bar{\ma{h}}}_{\text{DL-TA}_{i,d}} = (1 - \frac{1}{\alpha})  \hat{\bar{\ma{h}}}_{\text{DL-TA}_{i - 1,d}} + \frac{1}{\alpha}  \hat{\bar{\ma{h}}}_{\text{LSTM-DPA}_{i,d}}.
     \label{eq: proposed5}
\end{equation}

In this paper we used a fixed $\alpha = 2$ for simplicity. Therefore, the {\ac{TA}} applied in~{\eqref{eq: proposed5}} degrades the AWGN noise power $\sigma^2$ iteratively within the received {\ac{OFDM}} frame according to the ratio
\begin{equation}
	\small
	\begin{split}
			{R}_{\text{DL-TA}_{q}} &= \left( \frac{1}{4} \right)^{(q-1)} + \sum_{j=2}^{q} \left( \frac{1}{4} \right)^{(q-j+1)}=\frac{4^{q-1} + 2}{3 \times 4^{q-1}}.
	\label{eq:noise_degradtion}
	\end{split}
\end{equation}
    
    ${R}_{\text{DL-TA}_{q}}$ denotes the AWGN noise power ratio of the estimated channel at the $q$-th estimated channel, where ${1 < q < I + 1}$ and ${{R}_{\text{DL-TA}_{1}} = 1}$ denotes the AWGN noise power ratio at $\hat{\tilde{\ma{h}}}_{\text{LS}}[k]$. The full derivation of~\eqref{eq:noise_degradtion} is provided in Appendix A. It can be seen from the derivation of ${R}_{\text{DL-TA}_{q}}$ that the
    noise power is decreasing iteratively over the received {\ac{OFDM}} frame, hence the SNR increase which leads to better overall performance.

\begin{table}
	\renewcommand{\arraystretch}{1.4}
	\centering
	\caption{Proposed LSTM parameters.}
	\label{tb:LSTM_params}
	\begin{tabular}{l|l}
				\hline
		(LSTM units; Hidden size) & (1;128)  \\ \hline
		Activation function              & ReLU ($y= \max(0,x)$)                     \\ \hline
		Number of epochs        & 500                                \\ \hline
		Training samples        & 16000                             \\ \hline
		Testing samples        & 2000                             \\ \hline
		Batch size          & 128                                    \\ \hline
		Optimizer       & ADAM                                       \\ \hline
		Loss function      & MSE                                     \\ \hline
		Learning rate        & 0.001                                 \\ \hline
	    Training SNR        & 40 dB                                 \\ \hline
	\end{tabular}
\end{table}

\begin{figure*}[t]
	\setlength{\abovecaptionskip}{6pt plus 3pt minus 2pt}
	\centering
	\includegraphics[width=2\columnwidth]{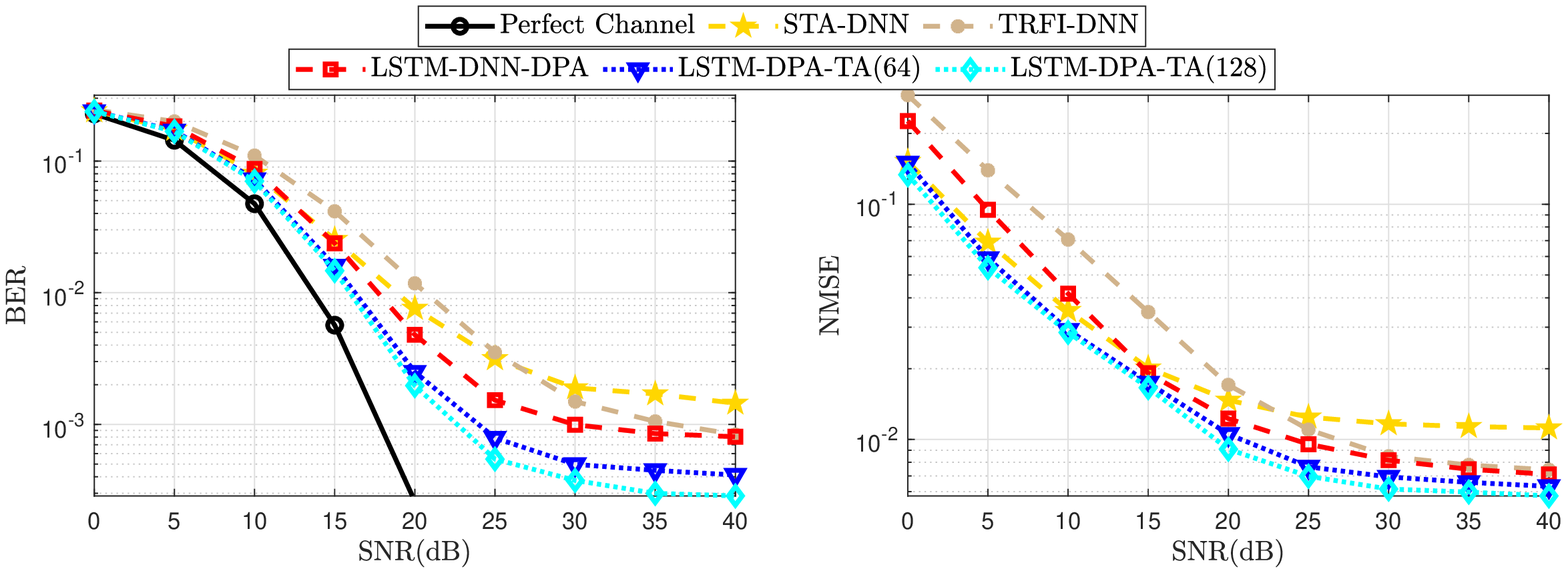}
	\subfloat[\label{High_BER_16QAM} BER performance employing high mobility scenario.]{\hspace{.5\linewidth}}
	\subfloat[\label{High_NMSE_16QAM} NMSE performance employing high mobility scenario.]{\hspace{.5\linewidth}} \\
	\vspace*{15pt}
	\includegraphics[width=2\columnwidth]{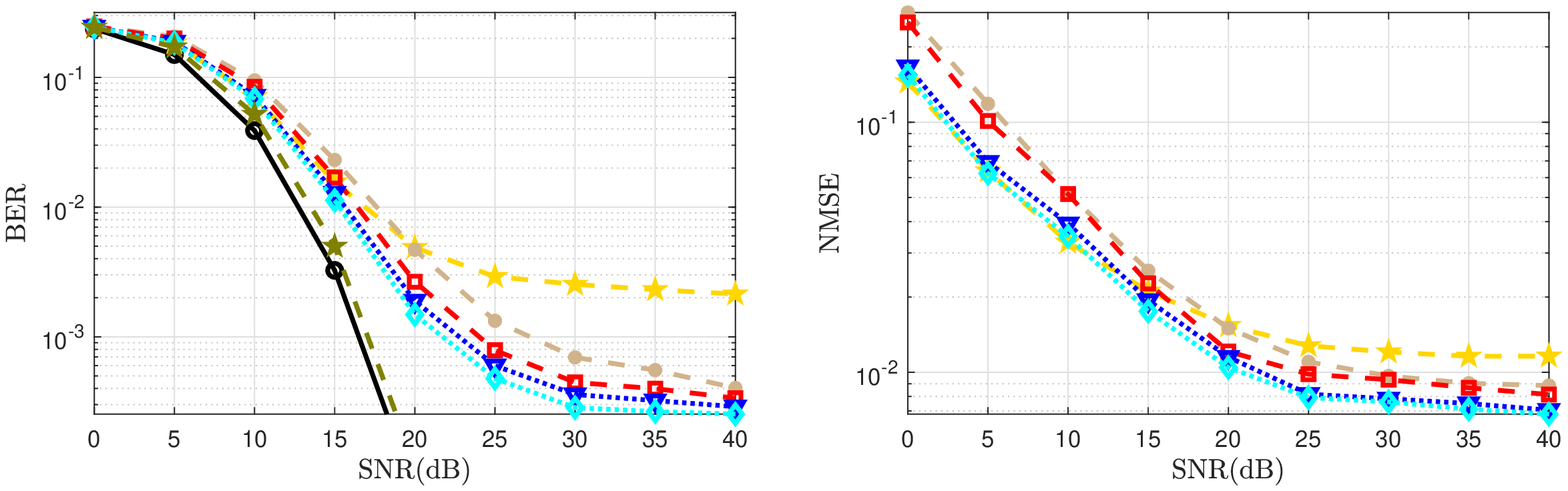}
	\subfloat[\label{Very_High_BER_16QAM} BER performance employing very high mobility scenario.]{\hspace{.5\linewidth}}
	\subfloat[\label{Very_High_NMSE_16QAM} NMSE performance employing very high mobility scenario.]{\hspace{.5\linewidth}}
	\caption{VTV-SDWW vehicular channel model simulation results.}
	\label{fig:High}
\end{figure*}

The proposed LSTM training is performed using {\ac{SNR}} = 40 dB to achieve the best performance as observed in~{\cite{r20}}, due to the fact that when the training is performed for a high {\ac{SNR}} value, the
LSTM is able to better learn the channel statistics, and due to its good generalization ability, it can still perform well in low {\ac{SNR}} regions, where the noise is dominant. Moreover, intensive experiments are performed using the grid search algorithm~{\cite{r210}} in order to select the best suitable LSTM hyper parameters in terms of both performance and complexity. Table.~{\ref{tb:LSTM_params}} summarizes the proposed LSTM training parameters.

%% file: simulation_results.tex
\section{Simulation Results} \label{simulation_results}



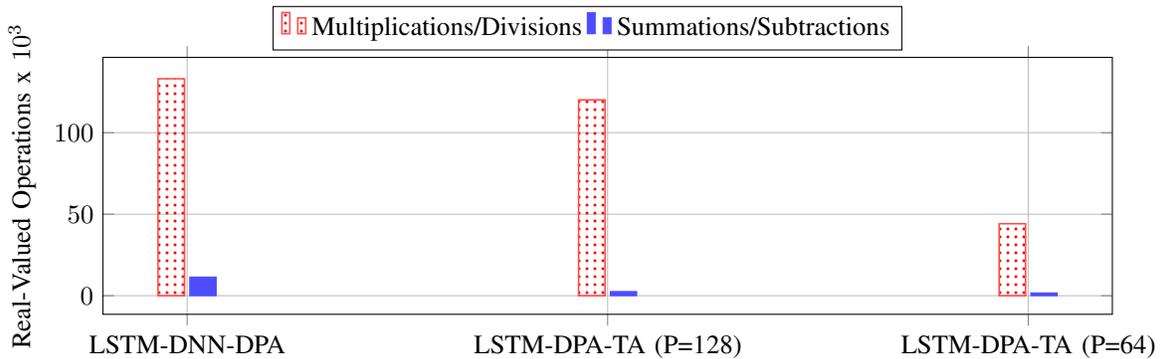
\begin {figure*}[t]
\centering
\begin{tikzpicture}
\begin{axis}[
    ybar,
    ylabel={Real-Valued Operations x $10^{3}$},
    symbolic x coords={LSTM-DNN-DPA, LSTM-DPA-TA (P=128), LSTM-DPA-TA (P=64)},
    xtick=data,
    legend style={at={(0.48,+1.2)},
    anchor=north,legend columns=-1},
    nodes near coords align={vertical},
    width=2\columnwidth,
    height=5cm, width=15cm,
    grid=major,
    cycle list = {red!70,blue!70,red!40,black!10}
    ]
\addplot+[semithick,
    pattern = dots,
    pattern color = red] coordinates {  (LSTM-DNN-DPA,133.088) (LSTM-DPA-TA (P=128), 120.136) (LSTM-DPA-TA (P=64), 44.168 )};

\addplot+[fill,text=black!10] coordinates { (LSTM-DNN-DPA,11.448)  (LSTM-DPA-TA (P=128), 2.560) (LSTM-DPA-TA (P=64),1.728 )};
\legend{Multiplications/Divisions, Summations/Subtractions}
\end{axis}
\end{tikzpicture}
\caption{Computational complexity of the studied channel estimators.}
\label{fig:bar_graph_LSTM}
\end{figure*}

In this section, {\ac{BER}} and  {\ac{NMSE}} simulations followed by a computational complexity analysis are conducted in order to evaluate the performance of the proposed estimator compared to IEEE 802.11p DL-based estimators.


In this paper, we consider the VTV-SDWW {\ac{TDL}} vehicular channel model~\cite{ref_CM} with two different mobility conditions: (\textit{i}) High mobility: V = 100 km/hr, with Doppler shift $f_{d}$ = 550 Hz. (\textit{ii}) Very high mobility: V = 200 km/hr, with $f_{d}$ = 1100 Hz. VTV-SDWW channel model represents the communication channel between two vehicles moving on a highway having center wall between its lanes, and it was obtained by a measurement campaign implemented in metropolitan Atlanta. Detailed measurement setups are provided in~\cite{ref_CMM}.
Concerning the simulation parameters, 16QAM modulation is employed over a frame size of $I = 50$ {\ac{OFDM}} symbols. The used channel coding is convolutional with half code rate. Moreover, three hidden layer DNN with $15$ neurons each is employed in the STA-DNN and TRFI-DNN estimators, with $\alpha = \beta = 2$ as defined in~\cite{ref_STA_DNN}. 
\subsection{BER and NMSE Performances}

Fig.~\ref{fig:High} depict the {\ac{BER}} and {\ac{NMSE}} performances of high and very high mobility vehicular scenarios respectively. As we can notice, the {\ac{DNN}}-based estimators suffer from error floor, especially in high {\ac{SNR}} region. Moreover, the {\ac{LSTM}}-{\ac{DNN}}-{\ac{DPA}} estimator outperforms them in the entire {\ac{SNR}} regions. This is explained by the high ability of the LSTM in capturing the temporal correlations of the channel more than the {\ac{DNN}} network.

On the other hand, the proposed {\ac{LSTM}}-{\ac{DPA}}-{\ac{TA}} estimator is able to outperform the {\ac{LSTM}}-{\ac{DNN}}-{\ac{DPA}} estimator in high mobility scenario by $7$ dB and $5$ dB gains in terms of {\ac{SNR}} at BER = $10^{-3}$, when $P = 128$ and $P = 64$ are employed respectively. We note that employing larger {\ac{LSTM}} hidden state size, i.e $P = 128$ achieves better performance than the optimized {\ac{LSTM}} unit ($P = 64$). However, in very high mobility scenario, since the temporal correlation between subsequent channel realizations significantly reduces, the proposed estimators outperform the {\ac{LSTM}}-{\ac{DNN}}-{\ac{DPA}} estimator by around $3$ dB gains in terms of {\ac{SNR}} at BER = $10^{-3}$. 
The proposed estimator performance gain in both scenarios is mainly due to employing the {\ac{TA}} processing which reduces the AWGN noise significantly. Moreover, the LSTM achieves better channel estimation performance than DNN while significantly reducing the computational complexity. This can be explained by the high ability of LSTM in learning the channel time correlations, compared with a simple DNN architecture.


 
\subsection{Computational Complexity Analysis}
In general, the estimator computational complexity is expressed in terms of real-valued multiplication/division and summation/subtraction  mathematical operations required to estimate the channel for one received {\ac{OFDM}} symbol. We note that {\ac{STA}}-{\ac{DNN}} and {\ac{TRFI}}-{\ac{DNN}} achieve lower complexity compared to the {\ac{LSTM}}-based estimators because {\ac{LSTM}} processing requires more operations than {\ac{DNN}} processing. Thus, in the following analysis we will focus on comparing the proposed {\ac{LSTM}}-{\ac{DPA}}-{\ac{TA}} estimator with the {\ac{LSTM}}-{\ac{DNN}}-{\ac{DPA}} estimator.

In~\cite{ref_STA_DNN}, the authors provide a detailed computational complexity analysis of the {\ac{DNN}}-based estimators, where the computational complexity of the {\ac{DNN}} network denoted by $C_{\text{DNN}}$ can be expressed as follows

\begin{equation}
\small
C_{\text{DNN}} = 2 \sum_{l=1}^{L+1} {N}_{l-1} {N}_l,
\label{eq:DNNcomp}
\end{equation}
where $L$ is the number of hidden layers within the {\ac{DNN}} network with $N_{l}$ neurons each, and $N_{0}$, $N_{L+1}$ denote the input and output {\ac{DNN}} network dimensions respectively. 

For the computational complexity of the  {\ac{LSTM}} unit, it can be calculated in terms of the required real values operations performed by its four gates, where each gate applies $P^{2} + P K_{in}$ real-valued multiplications, and $3P +  K_{in} - 2$ real-valued summations. In addition to $3P$  real-valued multiplications, and $P$ real-valued summations required by~\eqref{eq: lstm_cell_state}, and~\eqref{eq: lstm_hidden_state}. Therefore, the overall computational complexity for the {\ac{LSTM}} becomes
\begin{equation}
\small
C_{\text{LSTM}} = 4 (P^{2} + P K_{in} + 3P + K_{in} -2) + 4P.
\label{eq:LSTMcomp}
\end{equation}

The {\ac{LSTM}}-{\ac{DNN}}-{\ac{DPA}} estimator employs one {\ac{LSTM}} unit with $P=128$ and $K_{in} = 112$, followed by one hidden layer {\ac{DNN}} network with $N_{1} = 40$ neurons. Finally, the {\ac{LSTM}}-{\ac{DNN}}-{\ac{DPA}} estimator applies the {\ac{DPA}} estimation which requires $18 K_{d}$ real-valued multiplication/division and $8 K_{d}$ real-valued summation/subtraction. Therefore, the overall computational complexity of the {\ac{LSTM}}-{\ac{DNN}}-{\ac{DPA}} estimator is $512 K_{\text{in}} + 98 K_{d} + 71040$ real-valued multiplication/division and $4 K_{\text{in}} + 88 K_{d} + 6776$ real-valued summation/subtraction.

For the proposed {\ac{LSTM}}-{\ac{DPA}}-{\ac{TA}} estimator, it employs one {\ac{LSTM}} unit with $P=128$ as {\ac{LSTM}}-{\ac{DNN}}-{\ac{DPA}} estimator, or $P = 64$ when the optimized {\ac{LSTM}} unit architecture is employed. Moreover, the proposed estimator uses $K_{in} = 2 K_{on}$, and applies {\ac{TA}} as a noise alleviation technique to the $\hat{\bar{\ma{h}}}_{\text{LSTM-DPA}_{i,d}}$ estimated channel, that requires only $2 K_{on}$ real-valued multiplication/division and $2 K_{on}$ real-valued summation/subtraction. As a results, the proposed {\ac{LSTM}}-{\ac{DPA}}-{\ac{TA}} estimator requires $4 P^{2} + P  (8 K_{on} + 3) + 18 K_{d} + 2 K_{on} $  real-valued multiplication/division and $13P + 10 K_{on} + 8 K_{d} - 8$ real-valued summation/subtraction.

Based on this analysis, the proposed estimator achieves less computational complexity compared to the {\ac{LSTM}}-{\ac{DNN}}-{\ac{DPA}} estimator. It records $9.73 \%$ and $77.63\%$ computational complexity decrease in the required real-valued multiplication/division and summation/subtraction respectively, when the {\ac{LSTM}} unit is employed with $P = 128$ hidden size. On the other hand, more complexity reduction can be achieved when the optimized {\ac{LSTM}} unit is used with $P = 64$ hidden size, where the proposed estimator is able to decrease the complexity of the required multiplication/division and summation/subtraction by $66.81\%$ and $84.90\%$ respectively. It is worth mentioning that replacing the {\ac{DNN}} network by the {\ac{TA}} processing in order to alleviate the AWGN noise is the main factor in decreasing the overall computational complexity, where the proposed estimator outperforms the {\ac{LSTM}}-{\ac{DNN}}-{\ac{DPA}} estimator while recording a significant computational complexity reduction. Fig.~\ref{fig:bar_graph_LSTM} shows a detailed computational analysis of the benchmarked estimators in terms of real-valued operations.



%% file: conclusions.tex
\section{Conclusion} \label{conclusions}
In this paper, we have investigated the channel estimation challenge in vehicular environments. This challenge arises from the doubly selective nature of the vehicular channel, especially in high mobility scenarios. The recently proposed {\ac{DL}}-based IEEE 802.11p estimators have been presented and their limitations have been discussed. In order to overcome these limitations, we have proposed an {\ac{LSTM}}-based estimator, that employs an {\ac{LSTM}} unit for channel estimation, and a {\ac{TA}} processing as a noise alleviation technique. Simulation results have
shown the performance superiority of the proposed {\ac{LSTM}}-{\ac{DPA}}-{\ac{TA}} estimator over the recently proposed {\ac{DL}}-based estimators, while recording a significant reduction in computational complexity.


%% file: appendixA.tex
\appendix
\small
\subsection{{\ac{TA}} noise power alleviation ratio}\label{sec:TAMDFT_Noise_Power_Derivation}

According to the {\ac{TA}} processing applied in~\eqref{eq: proposed5}, and assuming that the noise terms of consecutive symbols are uncorrelated, the noise power for the $q$-th estimated channel is expressed as follows
\begin{equation}
    \small 
        {R}_{\text{DL-TA}_{q}} = \Ex{ \norm{\frac{\tilde{\ma{v}}_{{q-1}}}{2}  +  \frac{\tilde{\ma{v}}_{{q}}}{2}}^2} 
        = \frac{\sigma_{{q-1}}^2}{4} + \frac{\sigma_{{q}}^2}{4}.
\end{equation}

Where $\tilde{\ma{v}}_{{q}}$ denotes the AWGN noise at the $q$-th received {\ac{OFDM}} symbol. We consider that $\hat{\bar{\ma{h}}}_{\text{DL-TA}_{1}}=  \hat{\tilde{\ma{h}}}_{\text{LS}}$, therefore the noise power of the first estimated channel is $\sigma^2$, and the noise power enhancement ratio for the successive estimated channels can be computed as follows

\begin{equation}
\small
{R}_{\text{DL-TA}_{q}} = \left\{
        \begin{array}{ll}        
            1 ,&\quad q = 1 \\\\
            \frac{1}{4} + \frac{1}{4}  =  \frac{1}{2},&\quad q = 2 \\\\
             \frac{\ma{R}_{\text{DL-TA}_{q-1}}}{4} + \frac{1}{4},&\quad 3 < q \leq I+1
        \end{array}\right.
\label{eq:Noise_P1}
\end{equation}

The generalization formula of~\eqref{eq:Noise_P1} can be written as a sequence where the first element ${R}_{\text{DL-TA}_{1}} = 1$ as follows

\begin{equation}
\small
\begin{split}
   {R}_{\text{DL-TA}_{q}} &= \frac{1}{4} {R}_{\text{DL-TA}_{q-1}} + \frac{1}{4}  
    = \frac{1}{4} \left( {R}_{\text{DL-TA}_{q-1}} + 1 \right) \\
    &= \frac{1}{4} \left( \frac{1}{4} {R}_{\text{DL-TA}_{q-2}} + \frac{1}{4} + 1 \right) \\
    &= \frac{1}{4} \left( \frac{1}{4}  \left( \frac{1}{4} R_{\text{DL-TA}_{q-3}} + \frac{1}{4} \right)  + \frac{1}{4} + 1 \right) \\
    &= \frac{1}{4} \left( \frac{1}{4^{(q-1)-1}} {{R}}_{\text{DL-TA}_{q-(q-1)}} + \frac{1}{4^{(q-1)-1}} + \dots + \frac{1}{4^{0}} \right) \\ 
    &= \frac{1}{4} \left( \frac{1}{4^{q-2}} {{R}}_{\text{DL-TA}_{1}} + \frac{1}{4^{q-2}} + \frac{1}{4^{q-3}} + \dots +  \frac{1}{4^{0}} \right) \\ 
    &= \frac{1}{4} \left( \frac{1}{4^{q-2}} + \frac{1}{4^{q-2}} + \frac{1}{4^{q-3}} + \dots +  \frac{1}{4^{0}} \right).
\end{split}
\label{eq:Noise_P2}
\end{equation}

The sequence derived in~\eqref{eq:Noise_P2} can be written as follows:

\begin{equation}
\small
{R}_{\text{DL-TA}_{q}} =  \frac{1}{4} \left( \frac{1}{4^{q-2}} + \sum_{j = 2}^{q} \left({\frac{1}{4}}\right)^{q-j} \right) .
\label{eq:Noise_P3}
\end{equation}

Let $j^{\prime} = q - j$, then~\eqref{eq:Noise_P3} can be written in terms of $j^{\prime}$ according to the summation of geometric sequence rule~\cite{ref_geo} such that

\begin{equation}
\small
{R}_{\text{DL-TA}_{q}} =  \frac{1}{4} \left( \frac{1}{4^{q-2}} + \sum_{j^{\prime} = 0}^{q-2} \left({\frac{1}{4}}\right)^{j^{\prime}} \right) 
= \frac{4^{q-1} + 2}{ 3 \times 4^{q-1}}.
\label{eq:Noise_P5}
\end{equation}

%% file: acknowledgment.tex
\section*{Acknowledgment}
Authors acknowledge the CY Initiative of Excellence for the support of the project through the ASIA Chair of Excellence Grant (PIA/ANR-16-IDEX-0008).